# Why Do Polls Fail?

# The Case of Four US Presidential Elections, Brexit, and Two India General Elections*

Andreas N. Philippou

**ABSTRACT:** One of the most widely known and important applications of probability and statistics is scientific polling to forecast election results. In 1936, Gallup predicted correctly the victory of Roosevelt over Landon in the US presidential election, using scientific sampling of a few thousand persons, whereas the Literary Digest failed using 2.4 million answers to 10 million mailed questionnaires to automobile and telephone owners. Since then, polls have grown to be a multibillion flourishing and very influential and important industry, spreading around the world. Polls have mostly been accurate in the US presidential elections, with a few exceptions. Their two most notable failures were their wrong predictions of the US 1948 and 2016 presidential elections. Most polls failed too in the 2016 UK Referendum, in the 2014 and 2019 India Lok Sabha elections, and in the US 2020 presidential election, even though in the latter three they did predict the winner. We discuss these polls in the present paper. The failure in 1948 was due to non-random sampling. In 2016 and 2020 it was mainly due to the problem of non-response, despite weighting adjustments, and possible biases of the pollsters. In 2014 and 2019 it was due to non-response and political biases of the polling agencies and news outlets that produced the polls.

Keywords: Central limit theorem, confidence interval, Gallup poll, Biden, Trump, Truman, Roosevelt.

## 1. Introduction and Summary

Let $p$ be the unknown proportion of members in a population who possess an attribute. If in a random sample of $n$ members from this population, $r$ members are found to possess this attribute, then $f = r/n$ is used to estimate $p$. Sometimes people write $p = r/n \pm E$, where E is the margin of error of the estimate. By the Central Limit Theorem, for large n,

$$P\left(f - z_{a/2}\sqrt{\frac{f(1-f)}{n}} \leq p \leq f + z_{a/2}\sqrt{\frac{f(1-f)}{n}}\right) \cong 1 - a,$$

where $P(Z \leq z_{a/2}) = 1 - a$ and $Z$ is distributed as $N(0,1)$ (see, e.g. [9, pp 186-187] and [16, pp 210-211]).

------------------

For $a = 0.5$, $z_{a/2} = 1,96$, and hence $p$ lies between the limits

$$f - 1.96\sqrt{f(1-f)/n} \text{ and } f + 1.96\sqrt{f(1-f)/n}$$

with probability $\cong 95\%$. Since $f = r/n$, it may be stated with approximate probability 95%, that

$$p = \text{r/n} \pm \text{E}, \quad \text{where } \text{E} = 1.96\sqrt{(r/n)(1-(r/n))/n} \,,$$

with $\text{E} \leq 3\%$ for $n = 1,068,$ and $\text{E} \leq 3,5\%$ for $n = 800$.

The above confidence interval for $p$, using simple random sampling, has been essentially the basis for polling after 1948, when the pollsters failed to predict Truman's victory using quota sampling, and decided to replace it by probability sampling.

Twelve years earlier, in 1936, Dr. George Gallup [2] (see, also, Squire [21] and Warren [22]) predicted correctly the victory of Roosevelt over Landon in the US presidential election, using only 50,000 responses to a "scientific sample", whereas the Literary Digest failed using 2.4 million answers to a "straw sample" of mailed questionnaires.

Since 1936, the polls have grown to be a flourishing and very influential and important multibillion industry, spreading around the world. They have mostly been accurate in the US presidential elections, with a few exceptions. Their two most notable failures were their wrong predictions of the US 1948 and 2016 presidential elections. The polls failed too in the 2016 UK Referendum, in the 2014 and 2019 India Lok Sabha elections, and in the US 2020 presidential election, even though in the latter three they did predict the winner.

The failure in 1948 was due to non-random sampling. In 2016 and 2020 it was mainly due to the problem of non-response. In 2014 and 2019 it was due to non-response and political biases of the polling agencies and news outlets that produced the polls.

For decades, polls were typically conducted by telephone, using live interviewers. Today, internet surveys, random digit dial (RDD), and Interactive Voice Response (IVR) polls are increasingly common. Most sampled persons, more than 90%, refuse to answer to polls. In particular, according to the Obama 2012 presidential campaign whiz-kid, David Shor [12, 13], in RDD polls roughly 1 percent of people respond. But those who respond to polls are "weird", they are not the same as those who do not, and this is biasing the polls.

## 2.   Roosevelt against Landon in 1936

In one form or another opinion polls have been part of the American scene for more than 150 years and slowly-slowly became more scientific and their use spread around most of the world.

During 1920-1932, the mass periodical Literary Digest became very famous for its successful predictions of the winner of the US presidential elections, based on very large samples of persons.

In 1936, however, the series of successes ended dramatically, and the Digest seized publishing in 1938. The periodical mailed 10 million "ballot-questionnaires" to automobile and telephone owners, and, on the basis of about 2.4 million "ballots" (2,376,523 to be exact) received back, it predicted a landslide victory 3 to 2 for Alfred Landon against Franklin Roosevelt. Landslide victory it was, but Roosevelt was the winner with 61%, and he became USA President, again, not Landon.

Two were the reasons for the failure of the Digest. First, the "straw sample" of 10 million recipients was not representative of the American voters, and, second and more important, the 2.4 million respondents were not representative of the 10 million recipients (the 6.4 million non-respondents were different). According to Squire [21], who used data from a 1937 Gallup survey which asked about participation in the Literary Digest poll, the magazine's sample and the response were both biased and jointly produced the wildly incorrect estimate of the vote. However, he states, if all of those who were sampled had responded, the magazine would have at least predicted Roosevelt as the winner. See also Bryson [2], who disputes the first reason altogether and writes about the making of a statistical myth. In contrast, Gallup, Roper and Crossley used "scientific sampling methods" designed to include the proper proportion of voters from each economic stratum - not just those who owned automobiles and phones, and their predictions, based on 50,000 responses, were closer to the actual landslide victory of Roosevelt. See [2], [21] and [22]. All three failed, however, 12 years later.

## 3. Truman against Dewey in 1948

Gallup, Roper and Crossley wrongly predicted Dewey's victory with 6, 15 and 5 percentage points more than Truman's, respectively, when in fact Truman won the US Presidency with a margin of 5 points and 114 Electors more than Dewey.

Table1. 1948 US Presidential Election

| Candidates | Gallup Poll | Roper Poll | Crossley Poll | Electors | Votes |
|------------|-------------|------------|---------------|----------|-------|
| Harry Truman | 44% | 38% | 45% | 303 | 50% |
| T. Dewey | 50% | 53% | 50% | 189 | 45% |
| S. Thurmond | 6% | 9% | 5% | 39 | 5% |

All three Pollsters used quota sampling in order to ensure that the sample represents the voters in various strata (residential area, sex, age, race, economic status), and interviewers. Each interviewer was assigned specified numbers in each stratum. The Chicago Tribune felt so confident in the polls that the night of the election went ahead and printed the following morning's edition with the headline DEWEY DEFEATS TRUMAN. The picture below, showing President Truman holding the paper, is one of the most famous images in American politics.

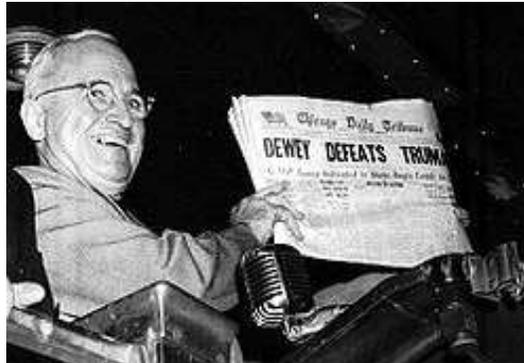

*Truman holding paper: Dewey Defeats Truman*

What went wrong is that each interviewer was free to pick the voters in each category anyway he pleased, not randomly. Evidently, more Republicans were interviewed than Democrats. In the aftermath, the three Pollsters and other members of AAPOR, which was founded in 1947, held a meeting in Iowa City and developed industry standards for public opinion polls, inaugurating a shift from quota sampling to probability sampling.

For decades, starting in the 1970's, polls were typically conducted by telephone, using live interviewers.

In 1980, the ASA Section on Survey Research Methods published the booklet What is A Survey? by R. Ferber et al. [5] (see, also, Fritz Scheuren [15] for an updated version of the same tittle) to help avoid mistakes and to promote a better understanding of what is involved in carrying out sample surveys correctly - especially those aspects that have to be taken into account in evaluating the results of surveys. The sample from the target population should be taken randomly, in order to be representative of the population, and the nonresponse should be avoided or minimized as much as possible.

Nowadays, internet surveys, random digit dial (RDD) telephone polls, and Interactive Voice Response (IVR) polls are increasingly common (see, for example, [8], [10] and [13]. But now more people refuse to respond, 90 to 99 percent, and this is biasing the polls despite weighting adjustments for education, age, race, gender, etc.

## 4. Trump against Clinton in 2016 and Trump against Biden in 2020

Donald Trump, a billionaire businessman and television star with a huge following on twitter, was a political outsider until he decided to run for US President in 2016. He defeated the Republican political establishment of 17 contenders, winning the Republican nomination, and faced Democrat Hillary Clinton, former First Lady and Secretary of State, who won the Democratic nomination against Bernie Sanders. During the campaign, almost every national poll and most state polls predicted Trump's defeat. Yet Trump won 306 electors to Clinton's 232, even though he lost the national popular vote by 2.1% and almost 3 million votes. He became president of the US receiving 304 electoral votes to Clinton's 227, as 7 faithless electors refused to vote for the one, they had pledged to vote for.

Table2. 2016 US Election Results

| Candidates | Vote Percentage | Electors | Number of Votes |
|---|---|---|---|
| Donald Trump | 46.1% | 306 | 62,984,828 |
| Hillary Clinton | 48.2% | 232 | 65,853,514 |

Trump's victory like that of Harry Truman in 1948, is considered as one of the greatest political upsets in modern U.S. history [6].

What went wrong with the polls? The polls did predict the winner of the national vote, albeit Clinton's recorded win margin of 2,1% was smaller than her predicted margin by most polls (4% by CNN Poll of Polls). But they failed to observe the swing to Trump of many white blue-collar workers within the Great Lakes and Rust Belt regions, to whom his rhetoric appealed, especially in Michigan, Pennsylvania and Wisconsin. The Latest Wisconsin polls | US Election 2016 poll tracker (ft.com), for example, on the day of the election found Clinton winning the state by a margin of 6.5%. Trump managed to win these States, and he won the election with 304 electoral votes.

According to Kennedy et al. [11], a committee, commissioned by AAPOR and headed by him, conducted an extensive investigation of the performance of pre - election polls in 2016. While the general public reaction was that the polls failed, the committee found the reality to be more complex. Some polls, indeed, had large problematic errors, but many polls did not. In particular, the national polls were generally correct (with respect to the popular vote) and accurate by historical standards. The most glaring problems occurred in state-level general election polling, particularly in the Upper Midwest.

The committee evaluated a number of different theories as to why so many polls underestimated support for Donald Trump. The explanations for which the most evidence exists are a late swing in vote preference toward Trump and a pervasive failure to adjust for overrepresentation of college graduates (who favored Clinton). In addition, there is clear evidence that voter turnout changed from 2012 to 2016 in ways that favored Trump. Despite widespread speculation, there is little evidence that socially desirable responding was a major contributor to poll error. If there was a Shy Trump effect on responses, it does not appear to have been particularly large.

One encouraging result from the historical analysis, they state, is that there is no systematic bias toward one major party or the other in US polling.

One broader question raised is whether the polling problems in 2016 could reoccur. The 2016 election featured a number of unusual circumstances that are perhaps unlikely to repeat (e.g., both major party candidates being historically unpopular, split outcomes in the popular vote and Electoral College, nearly 14 million votes across three states breaking for a candidate by about 0.5%), but several structural weaknesses of polls are likely to persist. Errors in state polls like those observed in 2016 are not uncommon, even though 2016 was a particularly bad election for state polls. Finally, a late swing in favor of one candidate (as appears to have occurred in 2016) is not something that pollsters can necessarily guard against, other than by polling closer to Election Day.

Although Clinton typically conceded defeat to Donald Trump, the Democrats opposed him fiercely, never accepting his victory. December 18, 2019, Donald Trump was impeached by the House of Representatives on charges of abuse of power and obstruction of Congress, but he was acquitted by the Senate on February 5, 2020.

In the US 2020 presidential election Biden defeated incumbent President Trump, as projected by almost all pollsters, even though the national election vote lead of 4,4% is quite smaller than the projected lead of most of them (10% by the CNN poll of polls as of 11/2/2020).

Table 3. 2020 US Election Results & Final CNN Poll of Polls

| Candidates | Votes | Poll of Polls | Electors | Votes |
|---|---|---|---|---|
| J. Biden | 51.3% | 52% | 306 | 81,268,924 |
| D. Trump | 46.9% | 42% | 232 | 74,216,154 |

The mistakes of the polls in Michigan, Pennsylvania and Wisconsin, underrepresenting Republicans, were even bigger.

Table 4. 2020 US Election Results & Final CNN Poll of Polls in Three States

| Candidates | MichVote | MichPoll | PennVote | PennPoll | WiscVote | WiscPoll |
|---|---|---|---|---|---|---|
| J. Biden | 50.6% | 51% | 50.0% | 50% | 49.4% | 52% |
| D. Trump | 47.8% | 42% | 48.8% | 44% | 48.8% | 42% |
| Biden's Lead | 2.8% | 9% | 1.2% | 6% | 0.6% | 10% |

Biden, however, managed to win back all three states plus Georgia, reaching 306 electors to Trump's 232, and became President-elect.

The U.S. Congress certified Biden's victory (of 306 Electors to 232) in the early hours of January 7, 2021, after supporters of President Trump, who attended his rally January 6, stormed the Capitol halting the certification process for several hours. As a result, 5 people died including a woman by gunshot, and a police officer. On January 13, 2021 the House of Representatives (all Democrats + ten Republicans) voted to impeach President Trump for incitement to insurrection. The Senate, however, acquitted him (57 guilty to 43 not guilty) on February 13, 2021.

Biden was inaugurated as the 46th President of the United States of America, Trump not attending, January 20, 2021. He leads a nation divided by the novel coronavirus pandemic, economic, racial and social turmoil, and fierce opposition from Trump and his followers.

What went wrong in the US 2016 and 2020 presidential election polls in the United States?

According to Kennedy et al. [11], national polls in 2016 were among the most accurate in estimating the popular vote since 1936. State-level polls, however, clearly under-estimated Trump's support in the Upper Midwest, due mainly to late change in vote preference and over-representation of college graduates.

The mistakes of the polls in 2020 have been bigger, underestimating Trump's national support and slightly overestimating Biden's, mainly due to the problem of non-response. But they correctly predicted the victory of Biden. According to a task force of AAPOR headed by Josh Clinton [3], after examining 2,858 national and state-level election polls, the polling error was of "unusual magnitude" resulting in the worst performance of national polls in 40 years and state polls in 20 years. Among polls conducted in the last two weeks before the election, the average signed error on the vote margin was too favorable for Biden by 3.9% in the national polls, and by 4.3% in state polls. See, also, Panagopoulos [13], who found that the pro-democratic bias of the polls in 2020 was systematic. According to David Shor [12, 13], a successful data scientist and 2012 Obama presidential campaign whiz-kid, every "high-quality public pollster" in the USA now does random digit dialing and roughly only 1 percent of people respond. Then, despite weighting for education, age, race, and gender, the pollsters fail, because respondents are quite different than nonrespondents.

## 5. 2016 UK Brexit Referendum

The United Kingdom joined the European Economic Community in 1973. On 23 June 2016, in a Referendum, the UK voted for *Leave* the European Union by 51.89% to 48.11% for Remain, a margin of 3.78%. However, even on the day of the Referendum a YouGov poll predicted 52% for *Remain* to 48% for *Leave*, using a sample of 4,772 voters!

The following day, the British Polling Council [1] reported its analysis of the final EU referendum polls. "Seven member companies issued 'final' polls of voting intentions in the EU referendum. While no company forecast the eventual result exactly, in three cases the result was within the poll's margin of error of plus or minus three points, and in one of them Leave was correctly estimated to be ahead. In the four remaining cases, however, support for Remain was clearly overestimated. This is obviously a disappointing result for the pollsters, and for the BPC, especially because every single poll, even those within sampling error, overstated the Remain vote share".

The error is mainly due to the problem of non-response. It is also likely that pollsters were biased in favor of Remain.

## 6. The 2014 and 2019 India Lok Sabha Elections

The 2014 India Lok Sabha election was held from 7 April to 12 May 2014. About 834 million people were eligible to vote, and turnout was over 66 per cent – the highest until then.

The National Democratic Alliance (Bharatiya Janata Party led) won 336 seats, the United Progressive Alliance (Indian National Congress led) won 59 seats, and Others won 149 seats. It was the greatest upset in Indian political history.

Almost all polls, including the exit polls, grossly underestimated the strength of NDA and overestimated the strength of the UPA, even though they did predict the victory of NDA.

Table 5. 2014 India Lok Sabha Election : Exit Polls

| Polling agency | NDA | UPA | OTHERS |
|---|---|---|---|
| CNN-IBN – CSDS – Lokniti | 276 ± 6 | 97 ± 5 | 148 ± 23 |
| India Today – Cicero | 272 ± 11 | 115 ± 5 | 156 ± 6 |
| News 24 – Chanakya | 340 ± 14 | 70 ± 9 | 133 ± 11 |
| India TV – CVoter | 289 | 101 | 148 |
| Poll of Polls | 283 | 105 | 149 |
| Election Results | 336 | 59 | 149 |

Jonah Force Hill [7] of Harvard, writing in the Diplomat ahead of the 2014 Lok Sabha (Lower House of Parliament) elections stated that, "Election polling in India is a notoriously unreliable exercise. It suffers from the political biases of the polling agencies and news outlets that produce the polls". He continued: "A more serious challenge to reliability comes from operational problems inherent in India's mammoth electorate, complex demographics, daunting geography and poor infrastructure, all of which make accurate polling an immensely labor intensive, expensive and often-dubious process". See, also Praveen Rai [16].

The 2019 India Lok Sabha election was held from 11 April to 19 May 2019. About 911 million people were eligible to vote, and turnout was over 67 per cent – the highest ever. The National Democratic Alliance won 353 seats, the United Progressive Alliance won 91 seats, and Others won 98 seats. Most of the polls in 2019, including the exit polls [12], underestimated NDA and overestimated UPA, even though they did predict the winner. However, the two exit polls [19, 20] shown in the following table, were almost excellent.

Table 6. 2019 India Lok Sabha Election: Two Successful Exit Polls

| Polling Agency | NDA | UPA | OTHERS |
|---|---|---|---|
| India Today-Axis My India | 352 ± 13 | 93 ± 15 | 82 ± 13 |
| News 24-Today's Chanakya | 350 ± 14 | 95 ± 9 | 97 ± 11 |
| Election Results | 353 | 91 | 98 |

## 7. Concluding Remarks

In ending the paper, we make the following remarks.

1. The correct prediction of President Roosevelt's reelection in 1936 by George Gallup, who used a small "scientific sample" of responding voters to interviewers, in contrast to the failure of the Literary Digest, which sent millions of "ballot-questionnaires" and received back as responding "ballots" only 25% of them, was the beginning of successful scientific polling in the United States, albeit with a few failures. Polls spread throughout most of the countries of the world.

2. The two most notable failures of pollsters in the US were their wrong predictions of the US 1948 and 2016 US Presidential Elections. In 1948, the failure was clearly due to non-random sampling. In 2016, it was mainly due to very high non-response percentage, 90 to 99, and the resulting bias despite weighting adjustments. It is true that the national polls did predict correctly the winner of the national vote. Several state polls, however, especially in Michigan, Pennsylvania and Wisconsin, failed to observe the swing to Trump of many white blue-collar workers. Even on the day of the election, Clinton was predicted to win Wisconsin by a margin of 6.5%! Trump won all these states, reaching 306 electors to Clinton's 232, and became President receiving 304 votes to Clinton's 227 as seven electors defected.

3. In the US 2020 Presidential Election, almost all pollsters correctly predicted Biden's victory over incumbent President Trump, even though the national election vote lead is quite smaller than the projected lead of most of them. After examining more than 2,800 national and state-level election polls, a task force of AAPOR found that the polling error was of "unusual magnitude", resulting in the worst performance of national polls in 40 years and state polls in 20 years. Among polls conducted in the last two weeks before the election, the average signed error on the vote margin was too favorable for Biden by 3.9% in the national polls, and by 4.3% in state polls. Panagopoulos [15] found that the pro-democratic bias of the polls in 2020 was systematic.

4. In the 2016 UK Brexit Referendum, the UK voted for Leave the European Union by a margin of 3.78%. However, even on the day of the Referendum a YouGov poll predicted a margin of 4% for Remain, using a sample of almost 5,000! As the British Polling Council reported, seven member companies issued 'final' polls of voting intentions in the EU referendum. In three cases the result was within the poll's margin of error of plus or minus three points. In the four remaining cases, however, support for Remain was clearly overestimated. This is obviously a disappointing result for the pollsters, especially because every single poll, even those within sampling error, overstated the Remain vote share. The error is mainly due to the problem of non-response. It is also likely that pollsters were not unbiased.

5. In the 1914 India Lok Sabha Elections, the National Democratic Alliance won 336 seats, the United Progressive Alliance won 59 seats, and Others won the remaining seats. Almost all polls, including exit polls, grossly underestimated the strength of NDA and overestimated the strength of UPA, mainly due to non-response, and political biases of the polling agencies and news outlets that produced the polls. In the 1919 India Lok Sabha Elections, the National Democratic Alliance won 353 seats, the United Progressive Alliance won 91 seats, and Others won the remaining seats. Although most of the polls and the exit polls underestimated the strength of UDA and overestimated the strength of UPA, they did predict the winner. Two of the exit polls were excellent.

The main problem in polling is the problem of non-response, despite weighting adjustments. The people who answer questions in polls are different from the kind of people who refuse to answer, and this is biasing the polls.

Department of Mathematics, University of Patras, Patras 26500, Greece. E-mail address: anphilip@math.upatras.gr /